\documentclass[aps,prb,twocolumn]{revtex4}

\pdfoutput=1

\usepackage{amssymb}
\usepackage{amsmath}
\usepackage{graphicx}
\usepackage{subfigure}
\usepackage{textcomp}
\usepackage{color}
\usepackage{amsfonts}
\usepackage{bbold}
\usepackage{dsfont}
\usepackage{epsfig}

\newcommand{\arctanh}[1]{\operatorname{arctan}}

\begin{document}



\title{Impurity-Ion pair induced high-temperature ferromagnetism in Co-doped ZnO}

\author{C.D. Pemmaraju$^a$, R.~Hanafin$^a$, T.~Archer$^a$, H.B.~Braun$^b$ and S.~Sanvito$^{a}$}
\affiliation{%
$^a$ School of Physics and CRANN, Trinity College, Dublin 2, Ireland
}%
\affiliation{%
$^b$ School of Physics, University College Dublin, Dublin 4, Ireland
}%

\date{\today}

\begin{abstract}
Magnetic $3d$-ions doped into wide-gap oxides show signatures of room temperature ferromagnetism, 
although their concentration is  two orders of magnitude smaller than that in conventional magnets. The 
prototype of these exceptional materials is Co-doped ZnO, for which an explanation of  
the room temperature ferromagnetism is still elusive. Here we demonstrate that magnetism 
originates from Co$^{2+}$ oxygen-vacancy pairs with a partially filled level  close to the ZnO conduction band 
minimum. The magnetic interaction between these pairs is sufficiently long-ranged to cause percolation at moderate 
concentrations. However, magnetically correlated clusters large enough to show hysteresis at room temperature 
already form below the percolation threshold and explain the current experimental findings. Our work demonstrates 
that the magnetism in ZnO:Co is entirely governed by intrinsic defects and a phase diagram is presented. This 
suggests a recipe for tailoring the magnetic properties of spintronics materials by controlling their intrinsic defects.
\end{abstract}

\maketitle

ZnO is a piezoelectric conductive oxide, in which free-carriers coexist with optical transparency \cite{ZnOpiezo,TCO}. 
If made magnetic, ZnO will become the ultimate multifunctional material, with semiconducting, magnetic, optical and 
mechanical properties. This will have a far reaching impact on the emerging field of spintronics \cite{spinbook}
with applications in optoelectronics \cite{opto} and quantum computing \cite{QC}. Moreover it will allow us to go 
beyond the (Ga,Mn)As paradigm \cite{GaMnAs}, whose practical use is severely hampered by the low ferromagnetic 
critical temperature. This is why ZnO:Co is perhaps the most studied among all the diluted magnetic oxides. 
Room-temperature ferromagnetism (RTF), first demonstrated by Ueda et al. \cite{ueda}, is now confirmed by a number 
of groups \cite{rode,Prellier,venky} (see Table I of the supplementary material). The experimental situation
is however still confused and here we list the main findings.

i) Spectroscopy confirms that Co$^{2+}$ substituting Zn is the center responsible for all the different magnetic phases 
found experimentally, including RTF \cite{Lee02,Tuan04,Koba05}, paramagnetism \cite{Wi04,Tuan04} and 
spin-glasses \cite{Peng05}. RTF is usually assigned from magnetometry \cite{ueda,rode,Prellier,venky}.

ii) The saturation magnetization $M_\mathrm{s}$ and the remanence are always small and secondary phases 
are often difficult to rule out. However, except for metallic Co, most of them are either non-magnetic or 
antiferromagnetic with low N\'eel temperatures (CoO, Co$_2$O$_3$, Co$_3$O$_4$, ZnCo$_2$O$_4$). 
The coercive field is typically small ($\sim$~100~Oe) and only weakly temperature dependent. 

iii) $M_\mathrm{s}$ is usually smaller than what is expected for Co$^{2+}$ with values as low as 
0.01~$\mu_\mathrm{B}$/Co \cite{Tuan04}, suggesting antiferromagnetic interaction among Co$^{2+}$ 
and frustration \cite{Wi04,Koba05,Risboud03}. 

iv) Growth conditions and annealing are crucial for the magnetic state. Chemical methods \cite{Wi04,Risboud03} 
and molecular beam epitaxy \cite{sati06} generally lead to paramagnetism, while pulsed laser deposition produces 
RTF films \cite{ueda,rode,Prellier,venky}. Typically oxygen deficient growth \cite{venky} at tuned substrate
temperatures \cite{Liu07,mike3,Dinia05} promotes RTF. Similarly, annealing in vacuum enhances 
the magnetic moment and produces ferromagnetism \cite{Tuan04,Dinia05,Cui06,Hsu06}, while annealing in
oxygen has the opposite effect \cite{Hsu06,Han05}. 

v) The role played by free carriers in establishing magnetism is unclear. Sequential annealing in reducing and 
oxidizing atmosphere reveals little correlation between the electrical conductivity and the magnetic state \cite{Khare06}. 
Similar conclusions are reached for Al- and H-doping \cite{Lee06}. 

vi) ZnO is often reported to be oxygen deficient. This has been attributed to either oxygen vacancies
(V$_\mathrm{O}$) \cite{ZnODefect,ZunDef} or multi-center H \cite{CVDW}, with Zn interstitials (Zn$_i$)
now ruled out by both experimental \cite{ZnODefect} and theoretical \cite{ZunDef,CVDW} 
evidence. Thus the promotion of RTF due to Zn vapour exposure \cite{Schw04} cannot be attributed to an increase of 
the Zn$_i$ concentration.

vii) Electron paramagnetic resonance \cite{EPR2} suggests the presence of two magnetic centers. These are both related 
to Co$^{2+}$, although they exhibit fine differences in the signal. Interestingly, for a nominal Co concentration about
5\% the two centers appear with similar abundance. 

Existing mechanisms for ferromagnetism in the diluted limit cannot explain this complex collection of phenomena. 
The $p$-$d$ Zener model \cite{Dietl} lacks of its foundations when applied to Zn$_{1-x}$Co$_x$O since there is little 
correlation between carriers and magnetism \cite{Khare06,Lee06}. When present, carriers are electrons and not holes
with small exchange coupling to the local spins and therefore the typical critical temperatures ($T_\mathrm{C}$) are 
tiny at realistic carrier densities. Similarly super-exchange must be ruled out \cite{GKRules}. This is short 
ranged and RTF can be obtained only for $x$ above the nearest neighbour (NN) percolation threshold.
For the wurtzite lattice this is 20~\%, much greater than the typical experimental concentrations. 

Finally a modification of the Zener scheme, called the ``donor impurity band exchange'' (DIBE), assumes that 
the magnetic interaction is mediated by large hydrogenic orbitals associated to intrinsic defects and predicts 
ferromagnetism below the donor percolation threshold, i.e. in absence of free carriers \cite{CoeyNM}.
Although the mean field $T_\mathrm{C}$ obtained with realistic parameters for Zn$_{1-x}$Co$_x$O is extremely 
small ($\sim$10~K) \cite{CoeyNM}, the model is frequently used to explain the experimental results. We have 
investigated such a model with Monte Carlo calculations (see supplementary materials) 
and demonstrated that for Co and donor concentrations respectively of 10~\% and 1~\%, the $T_\mathrm{C}$ is 
only a few degrees K. RTF is obtained only with unrealistically high values of the exchange coupling,
and therefore the DIBE scheme must also be ruled out. 

In the absence of any simple-scheme for ferromagnetism we turn to atomistic density functional theory (DFT). 
For this problem the standard approximations to the exchange and correlation potential (LDA and GGA) 
are not appropriate since they over-delocalize and under-bind the Co-$d$ shell resulting in their incorrect 
positioning with respect to the Fermi level, $E_\mathrm{F}$. These failures are only minor in the case of 
GaAs:Mn \cite{Gosia}, but they become a serious drawback for the oxides where the Co $d$ density of states 
(DOS) has important contributions in the ZnO band-gap. We therefore use an approximated version of the self-interaction
correction (ASIC) scheme \cite{DasSIC}, which is free from these problems and produces exchange parameters 
for transition metal monoxides superior to those obtained with LDA/GGA \cite{Akin}.

\begin{figure}[ht]
\epsfxsize=7.0cm
\centerline{\epsffile{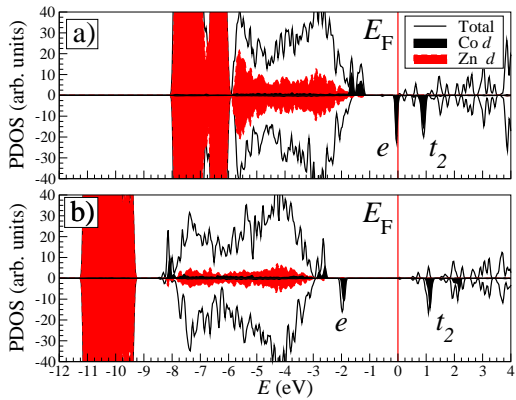}}
\caption{Density of states of Zn$_{1-x}$Co$_x$O as calculated from density functional theory. The simulation cell
is a 128 ZnO supercell in which one Zn atom is replaced by Co ($x$=0.0156). Panels (a) and (b) show
LDA and ASIC results respectively.}
\label{Fig1}
\end{figure}

In Fig.~\ref{Fig1} we present  the DOS of a Co impurity (at the Zn site) in a 128 atom ZnO supercell 
($x$=0.0156) calculated with both LDA and ASIC. Although they both predict a $2+$ valence, 
the position of the Co $d$ levels is remarkably different in the two cases.
LDA places the occupied minority $e$ states just below $E_\mathrm{F}$
at the edge of the ZnO conduction band minimum (CBM). ASIC shifts these by about 2~eV down to
the valence band top (VBT) in agreement with recent calculations \cite{Kata1}. In addition the 
ZnO band-gap opens and the Zn $d$ DOS
is also downshifted. This gives us a picture where there is no Co-$d$ contribution to the DOS around 
$E_\mathrm{F}$, with the first unoccupied minority states ($t_2$) placed at about 1~eV above the
CBM. Such a DOS is in excellent agreement with ultraviolet photoemission (UPS), which 
places the Co $d$ DOS at a binding energy of about 3~eV with a satellite peak at 7~eV and a rather
diffuse tail \cite{Koba05}. This electronic structure is incompatible with a carrier mediated picture of ferromagnetism.
In fact, $E_\mathrm{F}$ should be moved by at least 1~eV in order to affect the Co valence and 
therefore to promote the charge transfer necessary for strong exchange coupling \cite{CoeyNM}, a task hardly achievable.

We test this conjecture by calculating the magnetic coupling between two Co$^{2+}$ at various distances d$_\mathrm{Co-Co}$. 
In table \ref{Tab1} we show the magnetic coupling energy $E_\mathrm{M}$ ($E_\mathrm{M}>0$ indicates ferromagnetic coupling), obtained as the total energy difference between the ferromagnetic (FM) and antiferromagnetic (AFM) configuration of a selection of supercells containing two Co$^{2+}$ and one intrinsic defect far from both the Co$^{2+}$.
\begin{table*}[htb]
\caption{\label{Tab1}Calculated magnetic energy $E_\mathrm{M}$ for various magnetic centers and different
dopants configuration. C1 and C2 are the two magnetic centers included in the simulation cell and their relative
concentration (concentration of each center), D is the dopant with its concentration, d$_\mathrm{C1-C2}$ in the
distance between the two center expressed both in \AA\ and in NN shells. For some of the NN complexes we 
present the geometrical configuration (after relaxation in the pictures below).}
\begin{ruledtabular}
\begin{tabular}{llcccccl}
C1 ($x$) & C2 ($x$) & D ($y$)  & d$_\mathrm{C1-C2}$ (\AA) & d$_\mathrm{C1-C2}$ (NN)  & Position D & $E_\mathrm{M}$ (meV) & Fig. \\ \hline\hline
Co (0.015)    &    Co (0.015)    &   --                                       &   3.19 & 1 &   --      &  -38  & 1 \\
Co (0.015)    &    Co (0.015)    &   --                                       &   3.11 & 1 &   --      &   62  & 2 \\
Co (0.015)    &    Co (0.015)    &   --                                       &   4.54 & 2 &   --      &   -1  &  - \\
Co (0.010)    &    Co (0.010)    &Zn$_i$ (0.010)  &   8.01 & 3 &   Far   &   -1  &  - \\
Co (0.010)    &    Co (0.010)    &Zn$_i^+$ (0.010)  &   8.01 & 3 &   Far   &    1  &  - \\
Co (0.010)    &    Co (0.010)    & H (0.010) &   8.01 & 3 &   Far   &    0  &  - \\
Co (0.010)    &    Co (0.010)    &V$_\mathrm{O}$ (0.010) &   8.01 & 3 &   Far   &   -1  &  - \\ \hline
Co (0.015)    &    Co (0.015)    &Zn$_i$ (0.015)                  &  3.180& 1 &   Near   &   629  &  3 \\ 
Co (0.015)    &    Co (0.015)    &Zn$_i$ (0.015)                  &  2.551& 1 &   Near   &   3  &  4 \\ 
Co (0.015)    &    Co (0.015)    &Zn$_i$ (0.015)                  &  2.914& 1 &   Near   &   512  &  5 \\ 
Co (0.015)    &    Co (0.015)    &Zn$_i$ (0.015)                  &  2.557& 1 &   Near   &   731  &  3 \\ 
Co (0.015)    &    Co (0.015)    &V$_\mathrm{O}$ (0.010) &  2.585& 1 &   Near   &   10  &  6 \\ 
Co (0.015)    &    Co (0.015)    &V$_\mathrm{O}$ (0.010) &  2.795& 1 &   Near   &   -103  &  - \\ 
Co (0.015)    &    Co (0.015)    &V$_\mathrm{O}$ (0.010) \& Zn$_i$ &  2.315& 1 &   Near   &   899  &  - \\ 
Co (0.015)    &    Co (0.015)    &H (0.010)                            &  3.829& 1 &   Near   &   12  &  7 \\ 
Co (0.015)    &    Co (0.015)    &H (0.010)                            &  2.713& 1 &   Near   &   296  &  8 \\ \hline
CoV (0.015) & CoV (0.015)    & --                         & 5.55    & 2 & --        &  -6  & - \\
CoV (0.015) & CoV (0.015)    & H (0.010)                  & 2.30    & 1 & Far       &  423 & - \\
CoV (0.015) & CoV (0.015)    & H (0.010)                  & 2.27    & 1 & Far       &  614 & - \\
CoV (0.015) & CoV (0.015)    & H (0.010)                  & 5.55    & 2 & Far       &  84  & - \\
CoV (0.015) & CoV (0.015)    & H (0.010)                  & 4.51    & 2 & Far       &   9  & - \\
CoV (0.015) & CoV (0.015)    & H (0.010)                  & 6.94    & 3 & Far       &  20  & - \\
\end{tabular}
\vspace{0.3cm}
\centerline{
\framebox{\epsfxsize=1.9cm
\epsffile{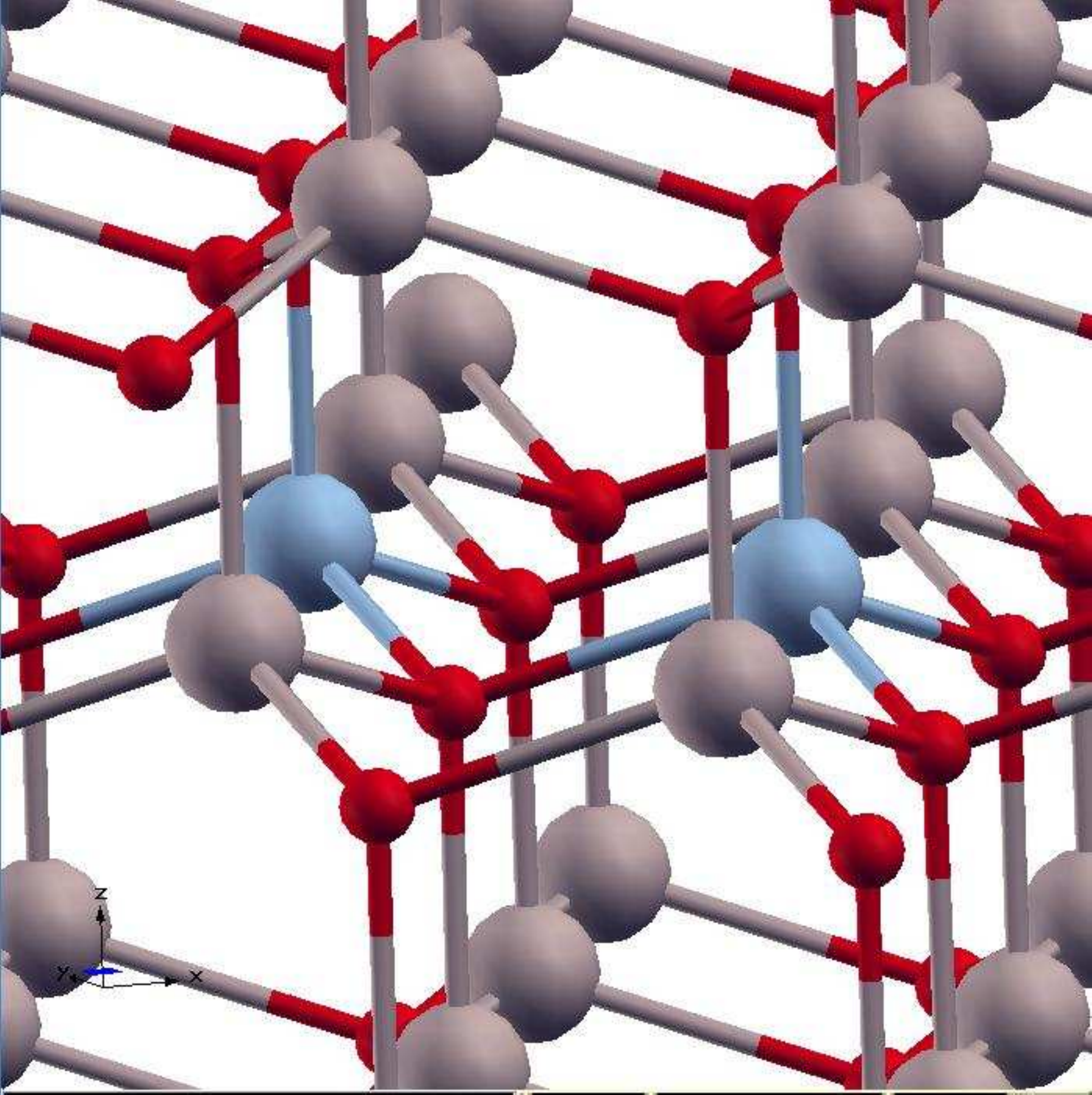}}
\framebox{\epsfxsize=1.9cm
\epsffile{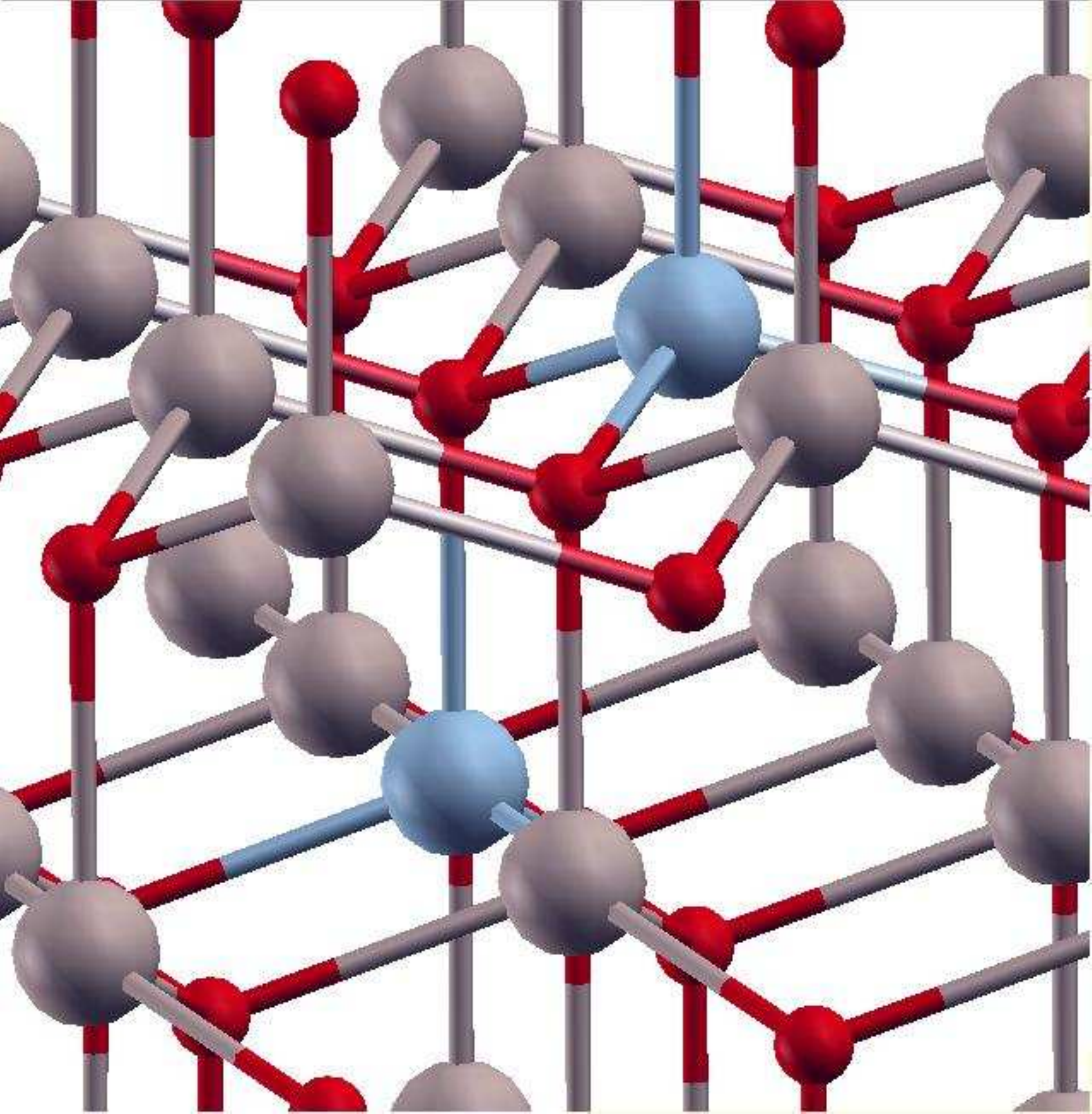}}
\framebox{\epsfxsize=1.9cm
\epsffile{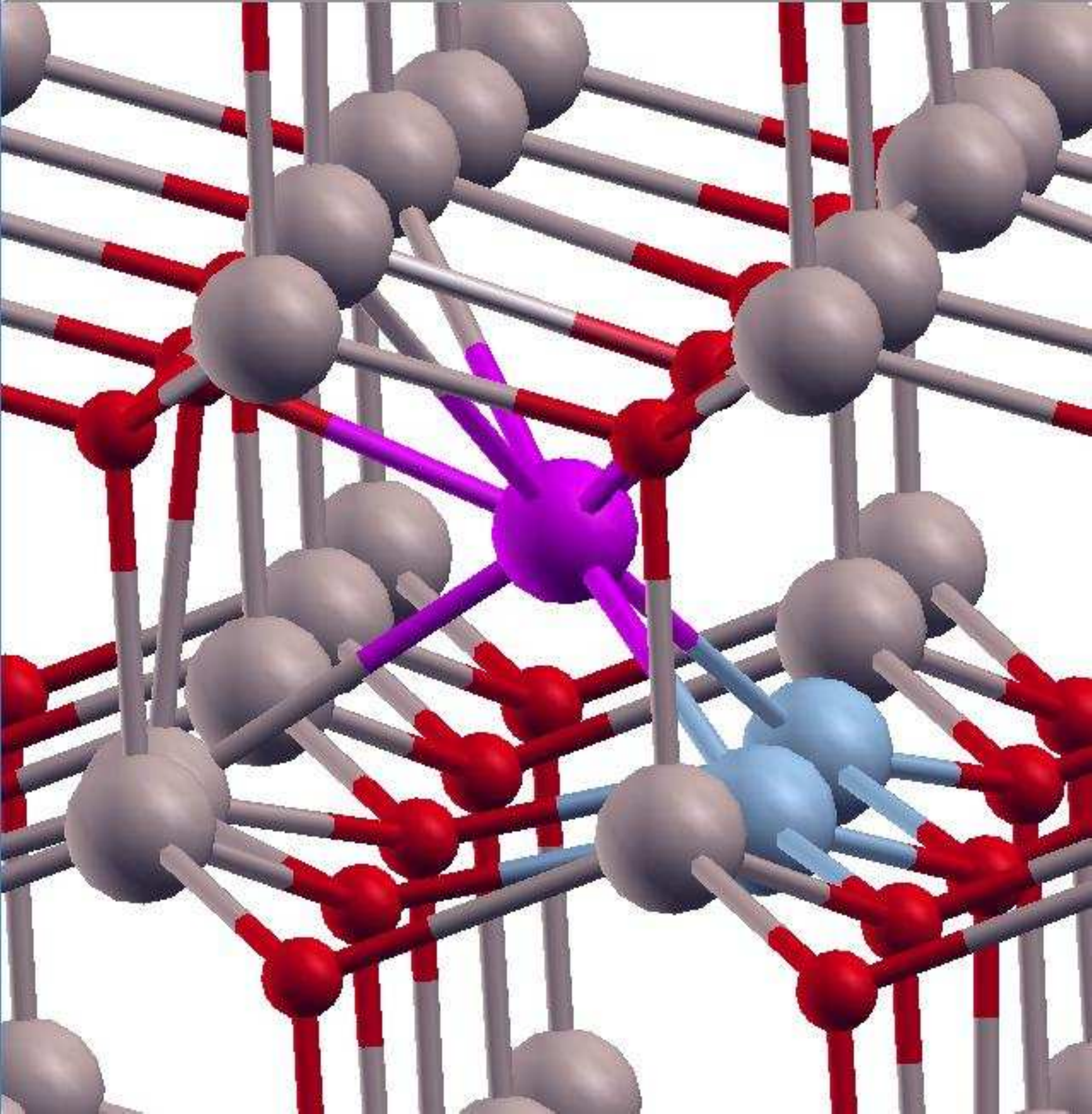}}
\framebox{\epsfxsize=1.9cm
\epsffile{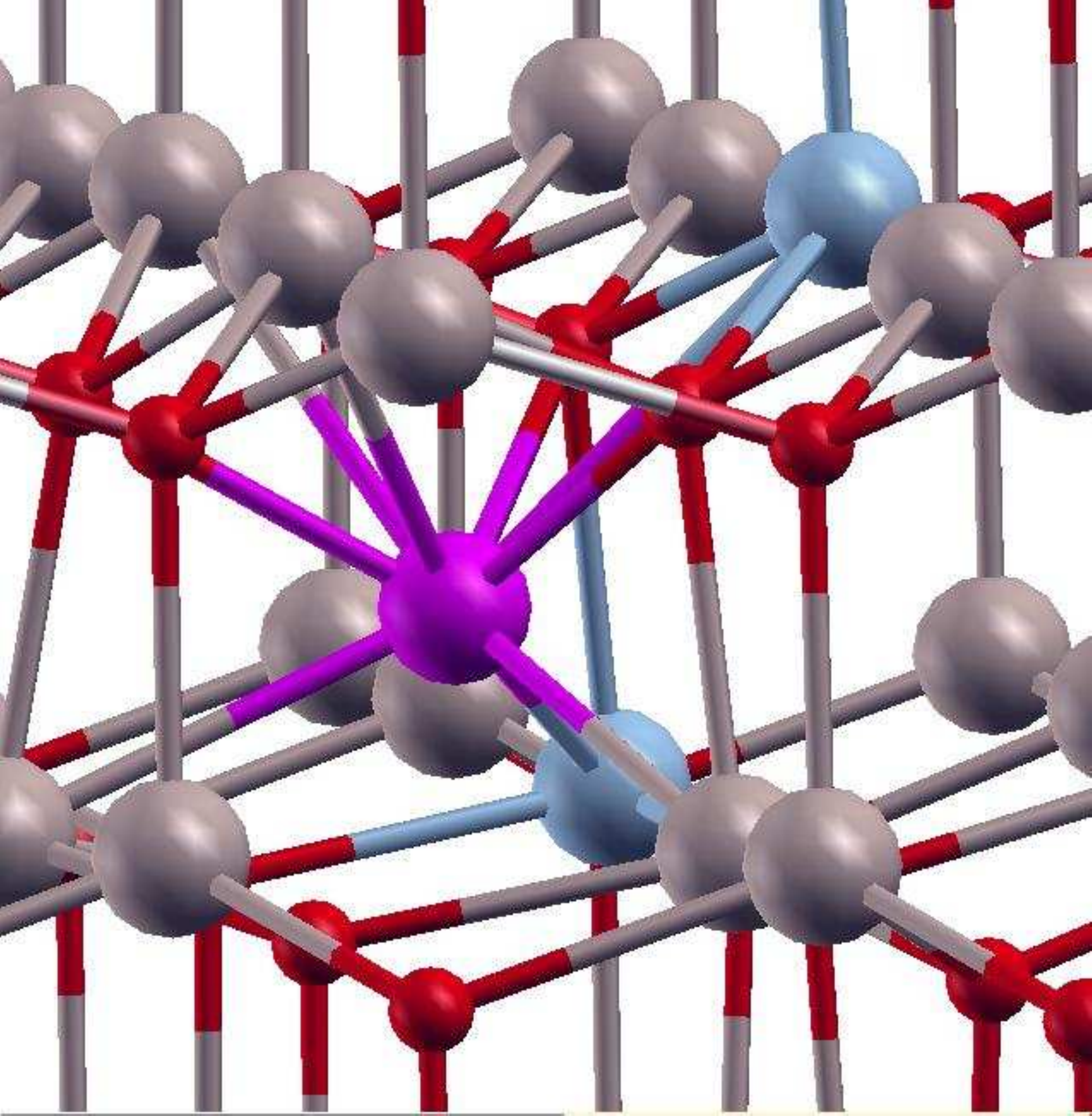}}
\framebox{\epsfxsize=1.9cm
\epsffile{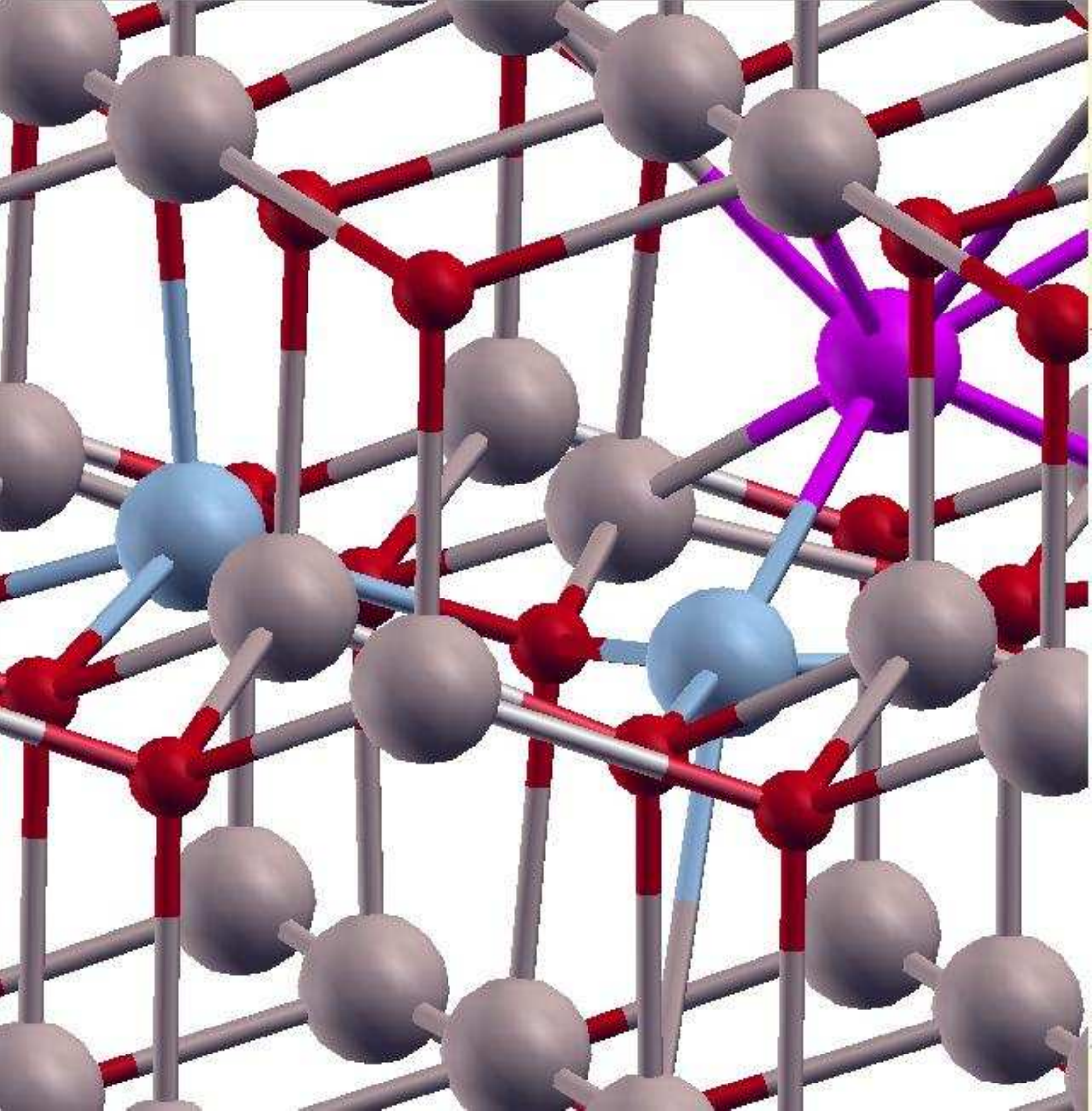}}
\framebox{\epsfxsize=1.9cm
\epsffile{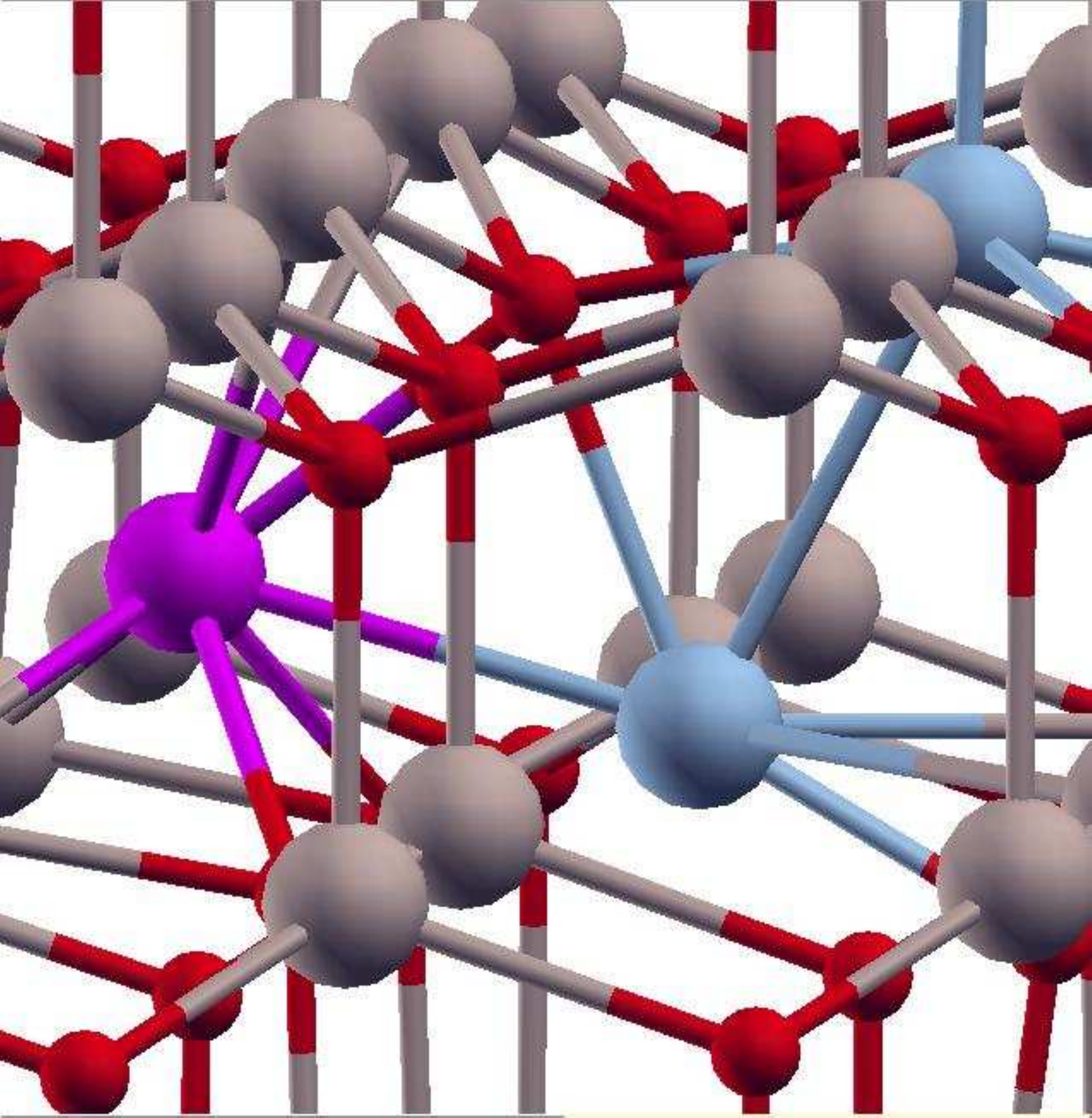}}
\framebox{\epsfxsize=1.90cm
\epsffile{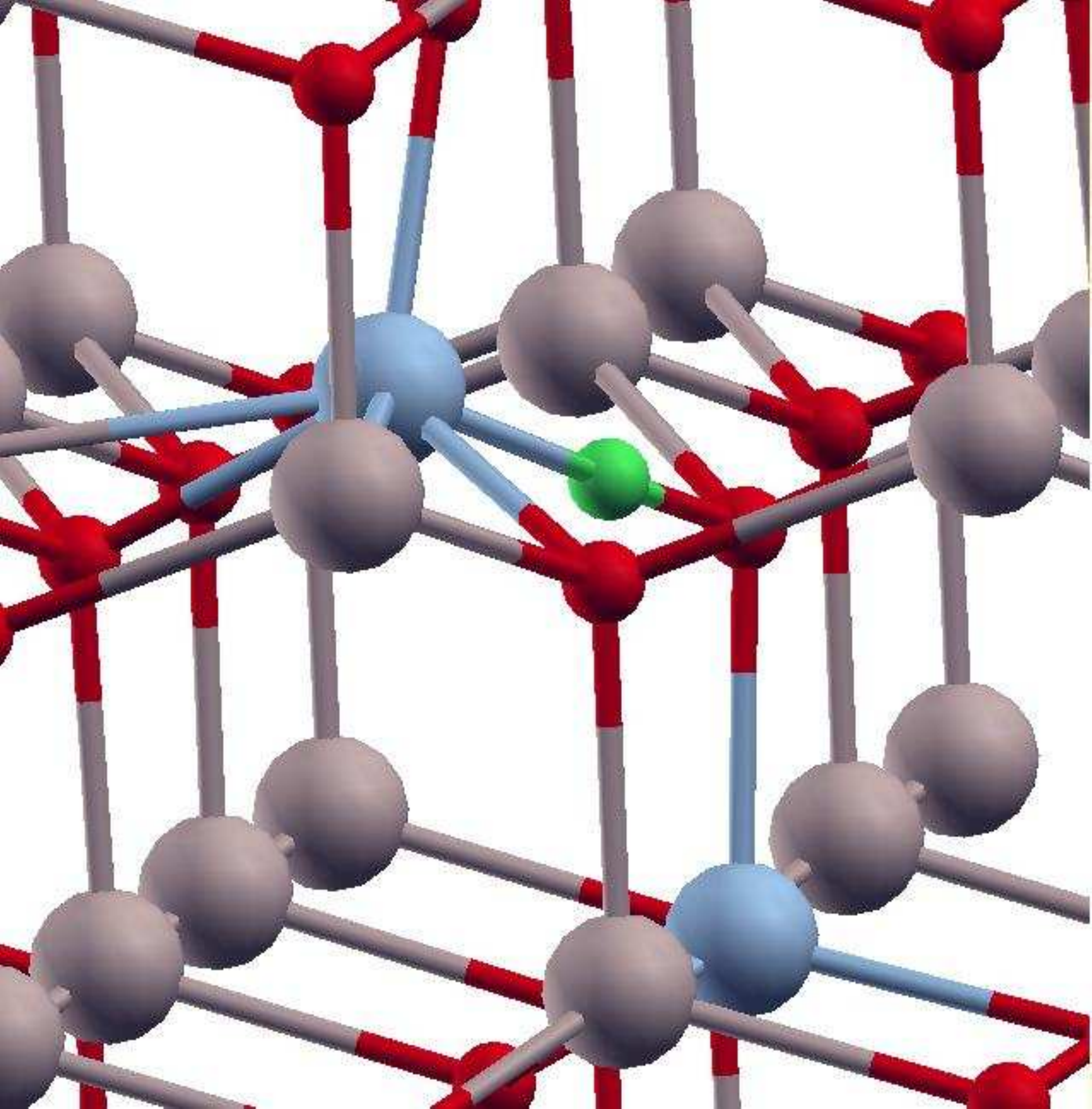}}
\framebox{\epsfxsize=1.9cm
\epsffile{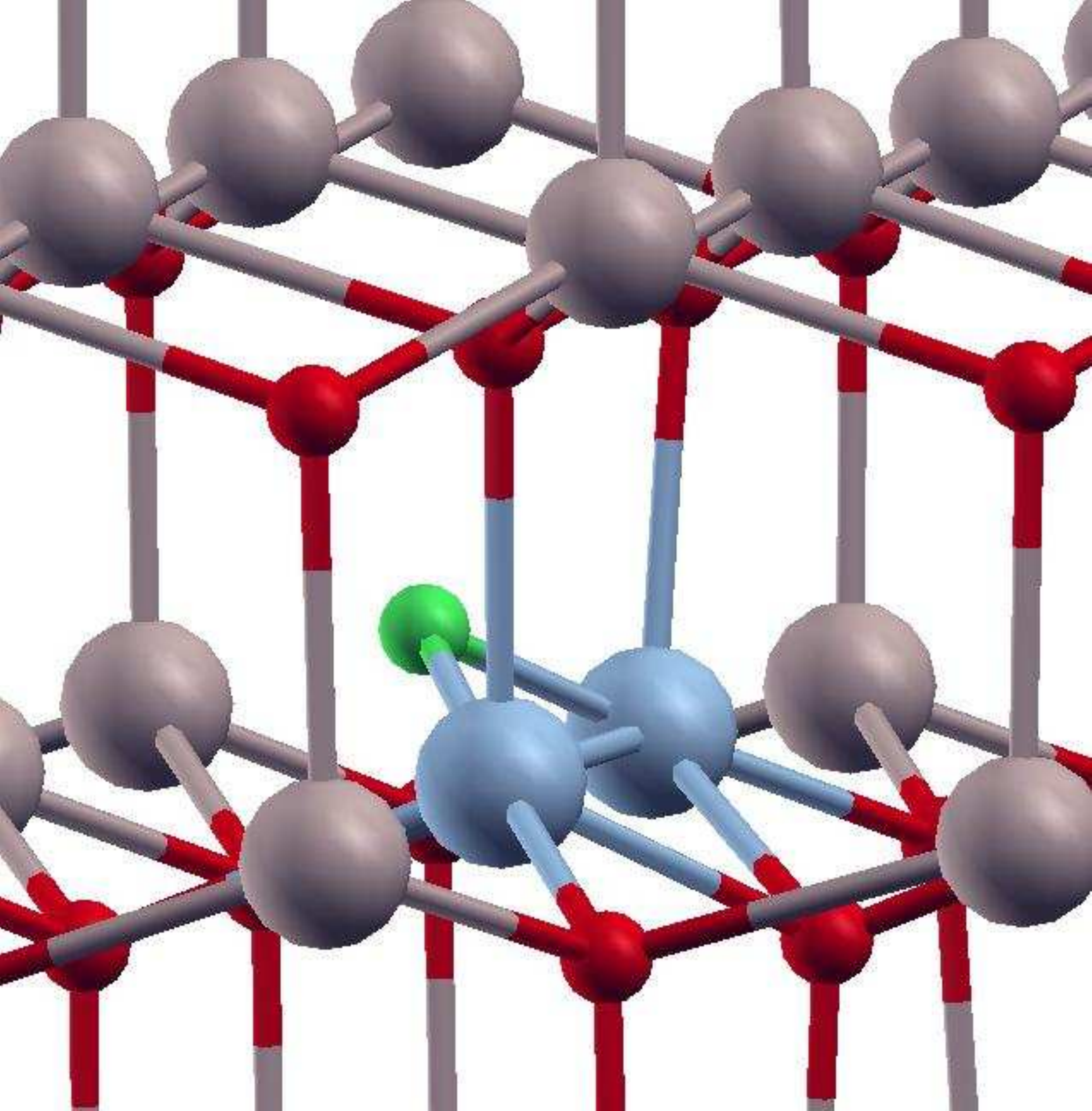}}
\epsfxsize=1.75cm
\epsffile{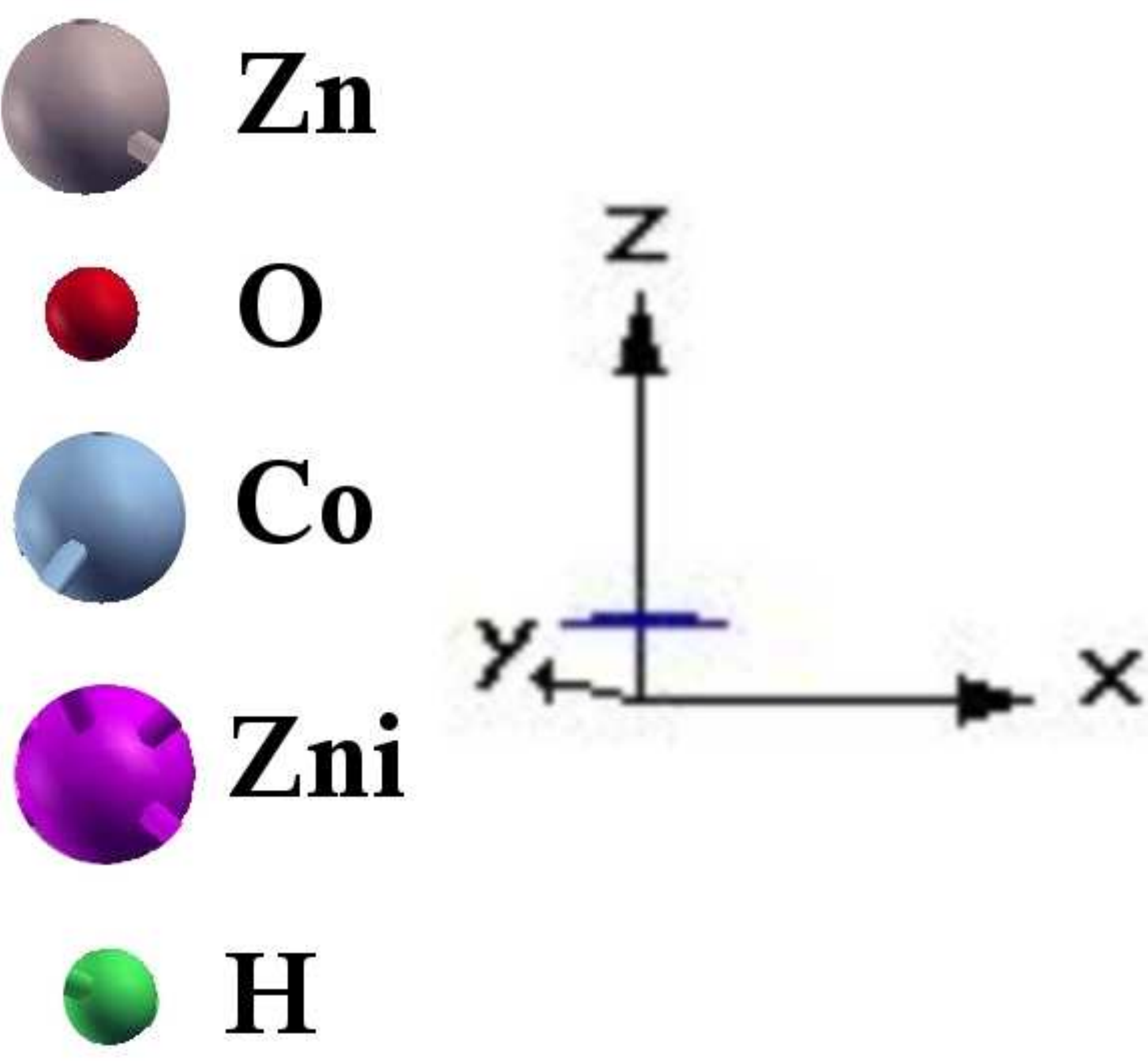}}
\end{ruledtabular}
\end{table*}
Clearly the exchange interaction between two Co$^{2+}$ is strong only at NN position, i.e. when the super-exchange 
interaction is effective. In this case the interaction is AFM in the $a$-$b$ plane (d$_\mathrm{Co-Co}$=3.19\AA) 
and FM along the $c$ axis (d$_\mathrm{Co-Co}$=3.11\AA). $E_\mathrm{M}$ however drops to zero 
already at second NN, regardless of the presence of additional intrinsic defects. In particular we show data for third NN (the data 
for second NN are similar) in presence of either Zn$_i$, H or V$_\mathrm{O}$ from which one has to conclude that RTF is not 
achievable by simply defect doping.

\begin{figure}[ht]
\epsfxsize=8.5cm
\centerline{\epsffile{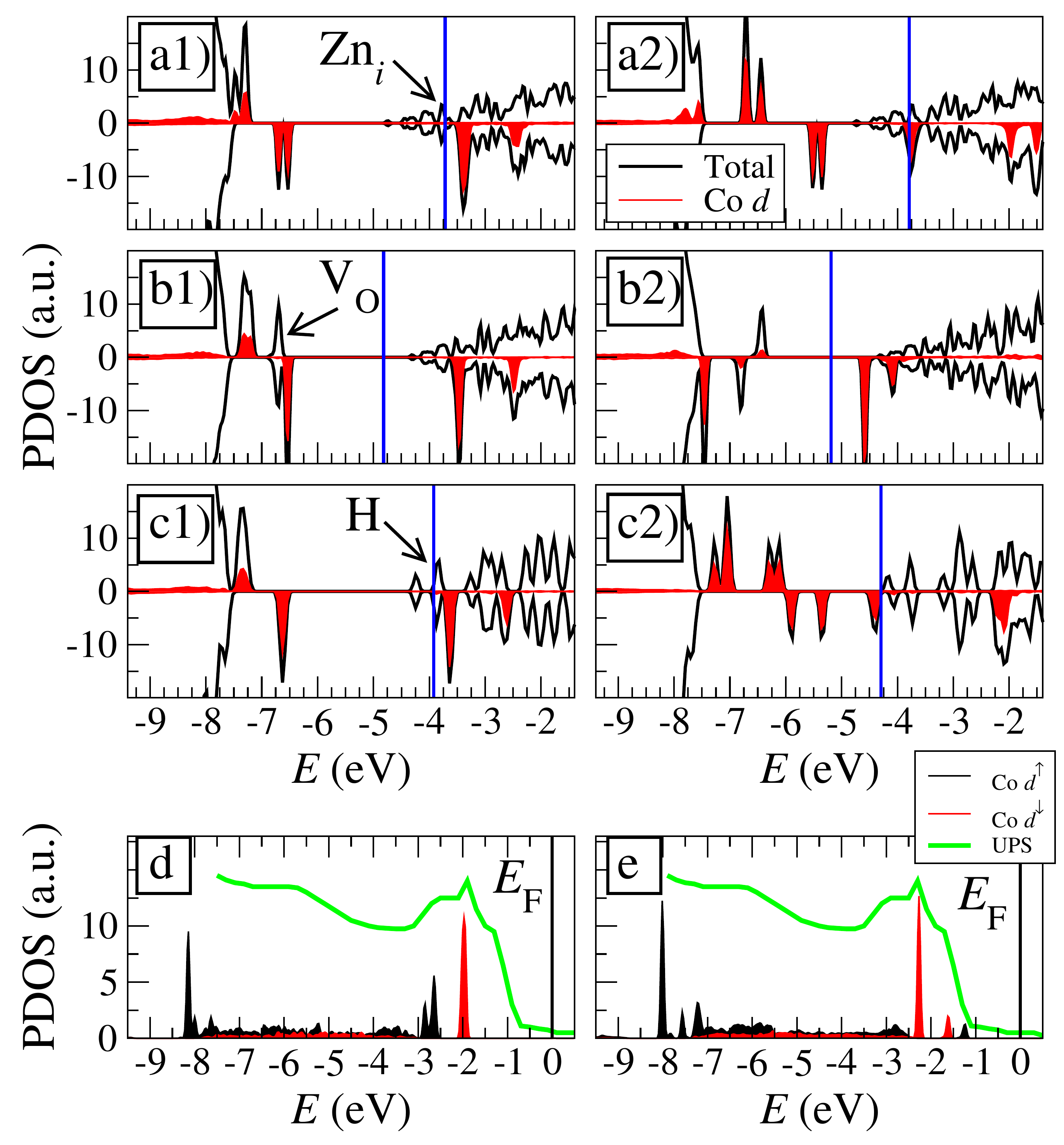}}
\caption{Density of states for a ZnO 128 atom supercell with one Co$^{2+}$ and one additional defect: a) 
Zn$_i$, b) V$_\mathrm{O}$, and c) H. The left panels are for Co$^{2+}$ and the defect well separated in the cell, while
the right panels are for the NN position. The arrows indicate the relevant defect position. In d) and e) we
present the Co 3 $d$ density of states as compared with UPS data from reference \cite{Koba05}. d) Co$^{2+}$
substitutional at the Zn site, e) Co$^{2+}$-V$_\mathrm{O}$ complex. The UPS signal has been aligned to the 
calculated DOS in order to have the first peak at the minority Co $e$ states.}
\label{Fig2}
\end{figure}

Figure \ref{Fig2} offers an insight on why Zn$_i$ and V$_\mathrm{O}$ are unable to mediate RTF. We present the DOS of a 
128 atom supercell containing one Co$^{2+}$ and one intrinsic defect (V$_\mathrm{O}$, Zn$_i$ and H), and compare 
the case where the Co$^{2+}$ and the defect are well separated in the cell with that in which they are at NN position.
In the case of distant defects the DOS is essentially a superposition of that of Co$^{2+}$ and the
defect. Both Zn$_i$ and H posses a filled hydrogenic level above the CBM and rather close to the
Co $d$-$t_2$ minority levels. In contrast V$_\mathrm{O}$ displays a doubly occupied
impurity level 1~eV above the VBT, almost at the same position of the minority Co-$d$ $e$ levels. 
Most importantly there is no evidence of interaction between Co$^{2+}$ and the defect levels (the magnetic 
moment calculated from the Mulliken population is $\sim$2.6~$\mu_\mathrm{B}$/Co for
all three cases, similarly to the case of Co$^{2+}$ only). This means that, despite the energy proximity of the 
defect levels to those of the Co 3$d$ shell, the overlap of the hydrogenic wave-function at the Co site is only 
minimal. For this reason we conclude that the DIBE model as formulated cannot be sustained by the electronic 
structure of Co$^{2+}$ in ZnO.

The right panels of Fig.~\ref{Fig2} for NN defects give a different picture. In the case of both H and Zn$_i$ there is
a substantial charge transfer from the defects to the Co$^{2+}$ resulting in a partial occupation of the
minority $t_2$ levels and a reduction of the magnetic moment (1.95~$\mu_\mathrm{B}$/Co and 
2.05~$\mu_\mathrm{B}$/Co respectively for Zn$_i$ and H). On-site repulsion moves upwards in energy
all the Co 3$d$ manifold and now the majority $t_2$ levels occupy the ZnO bandgap. In contrast, for V$_\mathrm{O}$ 
slight charge transfer from the Co 3$d$ to the vacancy and hybridization move the 3$d$ manifold downwards in energy and the 
doubly occupied V$_\mathrm{O}$ level spin-splits in the opposite direction to that of Co. 
Such a feature suggests that the exchange interaction between NN Co ions, when mediated by a defect, can be 
extremely large. This is indeed the case as demonstrated in table \ref{Tab1} where $E_\mathrm{M}$ for various 
Co-Co-defect complexes is presented. Such complexes effectively behave as small metallic clusters and it is of no 
surprise that the exchange energy increases quite dramatically for d$_\mathrm{Co-Co}$ around and below 2.5~\AA, 
which is the NN distance in metallic Co. 

Is this sufficient for RTF in the diluted phase? Unfortunately not. These interactions, although strong, are still short-range 
and therefore produce RTF only for $x$ around the percolation limit. Moreover, the strong FM interactions require a ratio 
between Co and the relevant donor of 2:1, which at percolation means typical defect concentrations of around 10\%. 
These are impossible to achieve for any of the defects investigated even under the most favorable conditions. Finally those 
complex structures, if abundant, should appear spectroscopically with a substantial increase of the Co-$d$ DOS in the 
ZnO bandgap. 

Since Co$^{2+}$ alone cannot be responsible for RTF at low dilution we have searched for other possible magnetic centers
and found that Co$^{2+}$-V$_\mathrm{O}$ pairs (CoV) overcome the limitations mentioned above. In figures \ref{Fig2}d 
and \ref{Fig2}e we present the DOS associated to the Co-$d$ shell for both Co$^{2+}$ and CoV, as compared with 
UPS from reference \cite{Koba05}. Clearly both Co$^{2+}$ and CoV are compatible with UPS,
in particular they both show a finite DOS at about -8~eV from $E_\mathrm{F}$. This feature is absent in the DOS
of both the Co-Zn$_i$ and Co-H complexes, which instead present substantial contributions in the ZnO gap and therefore 
are incompatible with the spectroscopy. Moreover CoV is the only complex among the ones studied which maintains 
Co in the 2+ valence state. 

CoV are also likely to be abundant. Our calculated V$_\mathrm{O}$ formation energy (see supplementary material)
in zinc rich conditions is 0.65~eV, which suggests that the V$_\mathrm{O}$ concentration in ZnO can be as 
large as 1~\% at equilibrium. Since Co does not introduce doping its presence will not change considerably the V$_\mathrm{O}$ 
formation energy. We have then only to establish whether the V$_\mathrm{O}$ sits preferentially close to a Co site. 
By using a 128 atom supercell we calculate a reduction in total energy of about 
340~meV for Co and V$_\mathrm{O}$ moving from third to first NN (pairing energy). This is quite a large gain suggesting 
that most of the V$_\mathrm{O}$ are indeed likely to be close to Co ions. The large pairing energy also means that the relative 
concentration of CoV ($x^\mathrm{CoV}$) with respect to that of Co$^{2+}$ ($x^\mathrm{Co}$) will 
increase upon oxygen absorbing processing. This is the case for both annealing in an oxygen poor atmospheres and
long exposure to Zn and Ti vapours \cite{ZnODefect}. In the first case one expects V$_\mathrm{O}$ migration to the Co sites, 
while in the second a preferential V$_\mathrm{O}$ formation close to Co. Finally we found that the pairing energy between two 
Co$^{2+}$ is also large ($\sim$~210~meV), while we do not find any substantial pairing interaction among the CoV 
in any charging configuration. We then expect an inhomogeneous distribution of the Co$^{2+}$, which in turn would lead
to an inhomegeneous distribution of CoV and the formation of both high and low concentration regions. 

Finally we have to establish whether CoV couple at long-range. Table \ref{Tab1} shows $E_\mathrm{M}$ for various CoV combinations.
Similarly to the case of Co$^{2+}$ also the magnetic coupling between two charge neutral CoV is remarkably weak already at second 
NN (-6~meV). However, when an additional electron donor is present, the situation changes dramatically with second 
and third NN magnetic coupling reaching up respectively to $E_\mathrm{M}\sim$80~meV and $E_\mathrm{M}\sim$20~meV. Note 
that we cannot exclude an even longer interaction range, which however is hardly accessible in our simulations since the size of 
the supercells becomes prohibitively large.

\begin{figure}[ht]
\epsfxsize=8.0cm
\centerline{\epsffile{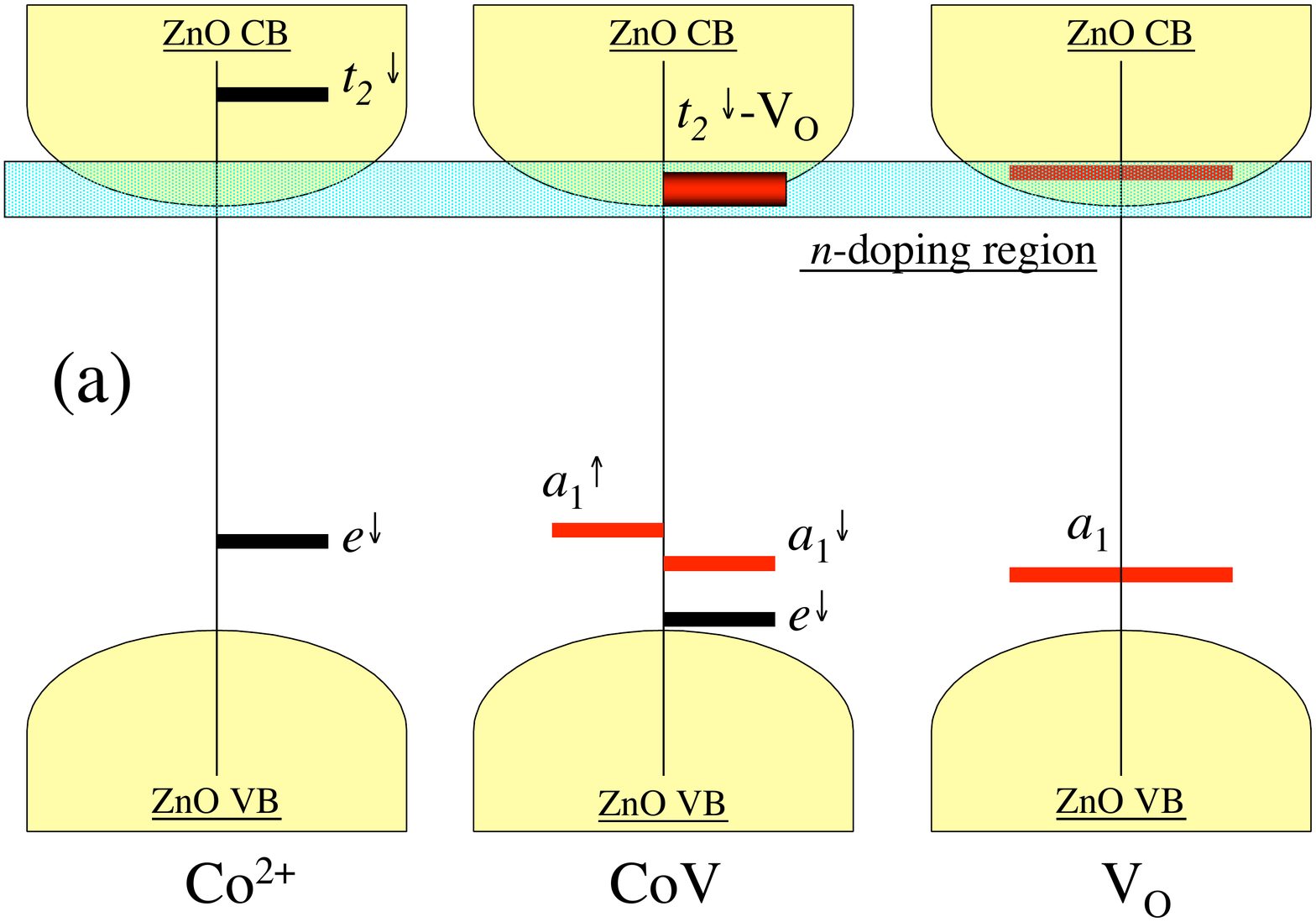}}
\epsfxsize=5.5cm
\centerline{\epsffile{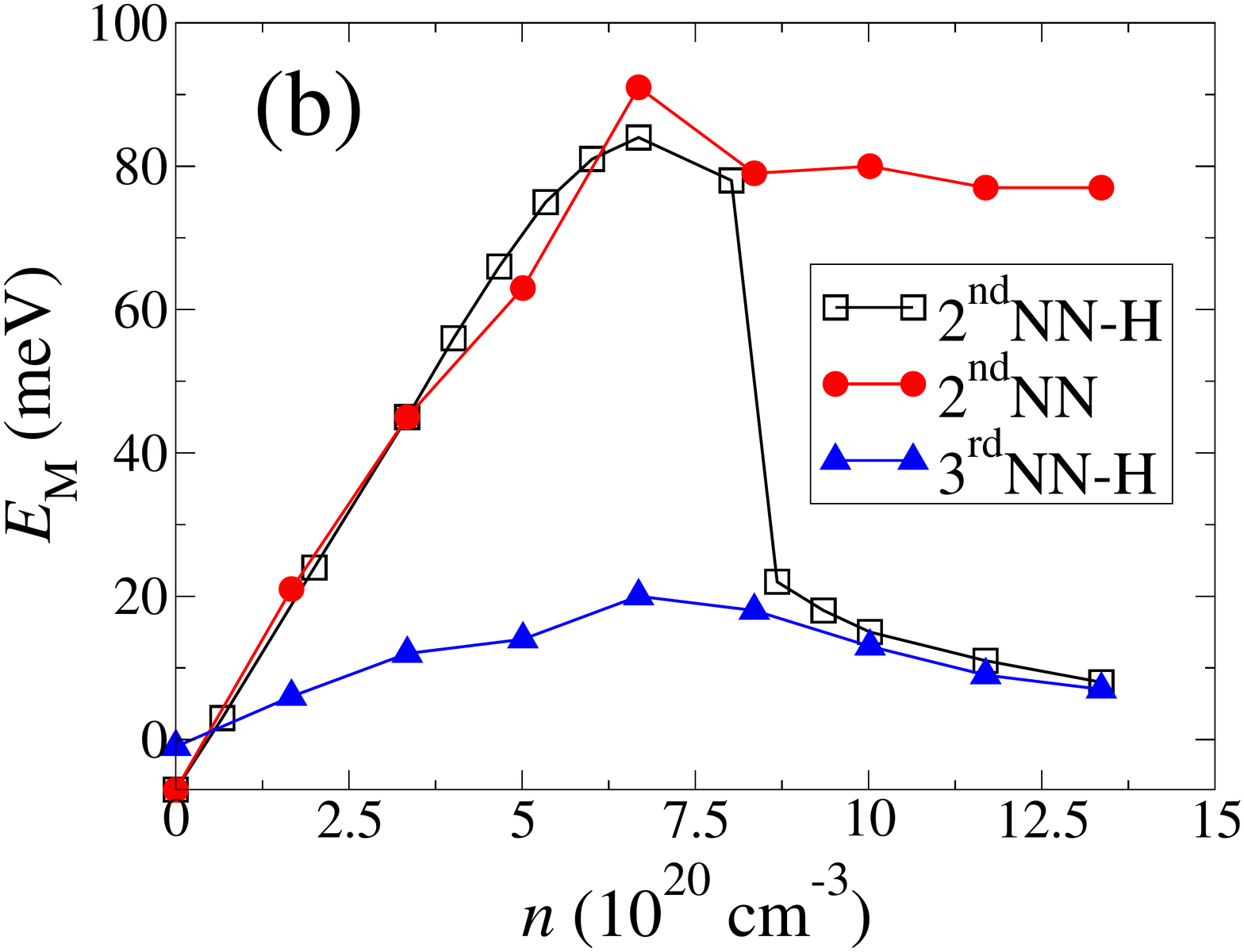}}
\caption{CoV impurity band. (a) level diagram for Co$^{2+}$, V$_\mathrm{O}$ and CoV, (b)
$E_\mathrm{M}$ for two 2$^\mathrm{nd}$ NN CoV as a function of the donor impurity band electron density. 
The electron density 6.7$\cdot10^{20}$~cm $^{-3}$ corresponds to one electron every two CoV.}
\label{Fig3}
\end{figure}
Why does CoV sustain long range coupling while Co$^{2+}$ does not? The cartoon of Fig.~\ref{Fig3}a shows
the electronic structure of Co$^{2+}$, V$_\mathrm{O}$ and CoV. The main feature is that the Co$^{2+}$ empty $t_2$ 
minority state broadens in CoV and forms a $t_2$-V$_\mathrm{O}$ hybrid level right at the CBM. This extends in space over the
the first shell of Zn ions and an impurity band forms already at tiny concentrations. Such a band can be easily $n$-doped.
Thus, while for Co$^{2+}$ the only exchange mechanisms available either involve virtual transitions or weak $s$-$d$ \cite{Lars} 
exchange interaction with hydrogenic orbitals, for CoV strong carrier-mediated magnetic interaction is possible. 
In figure \ref{Fig3}b we show $E_\mathrm{M}$ for two CoV in a 128 atom unit cell as a function of the electron doping. This is introduced by 
either moving $E_\mathrm{F}$ in our simulations or by explicitly introducing an H ion in the supercell. Clearly $E_\mathrm{M}$ depends 
strongly on the occupation of the impurity band, with a maximum at half filling ($n$=6.7$\times10^{20}$~cm$^{-3}$) and vanishing for $n$=0 
(empty band). 

Thus DFT offers us a mechanism for the ferromagnetism in (Zn,Co)O based on two magnetic centers. 
Co$^{2+}$ is responsible only for short range coupling, while CoV can instead sustain long range interaction via a 
fractionally filled impurity band. Can this alone produce RTF? Percolation theory \cite{percolation} sets a strict condition 
for a magnetic ground state of diluted systems: the concentration of magnetic impurity should exceed the percolation 
threshold $x_\mathrm{c}$. This depends on the range of the interaction and for the fcc lattice we find 19.8\%, 13.7\% and 6.2\%
for interaction extending respectively to 1$^\mathrm{st}$, 2$^\mathrm{nd}$ and 3$^\mathrm{rd}$ NN \cite{ZungerPer}. Therefore 
our two-center model already produces long range ferromagnetism at $x^\mathrm{CoV}> x_\mathrm{c}\sim 6\%$. However, we do not need such a 
large $x$ for observing hysteresis in the $M$-$H$ curve at room temperature (i.e. for explaining the various experimental claims). 
This can be achieved below $x_\mathrm{c}$ since one just needs a number of percolating clusters large enough to be 
superparamagnetically blocked. Note that ``clusters'' here mean regions where $x$ is locally larger than $x_\mathrm{c}$. 
The presence of CoV pushes this limit far below the 20\% needed by Co$^{2+}$ and by the recently proposed model
where the magnetism originates from uncompensated spins at the surface of antiferromagnetic clusters \cite{TDC}.

The size of those clusters can be estimated by considering coherent rotation of the magnetization over an anisotropy barrier 
$DN_\mathrm{B}S^2$ ($D$ is the zero-field splitting, $N_\mathrm{B}$ the number of magnetic ions magnetically blocked and $S$ 
the Co spin). By taking $D=2.76$~cm$^{-1}$ from EPR measurements \cite{sati06} we obtaine the estimate 
$N_\mathrm{B}\sim800$ for a blocking temperature $T_\mathrm{B}=300$~K. This however is rather conservative. 
In granular magnets random dipolar interaction \cite{Allia}, random magnetic anisotropy \cite{Beni} or spinodal 
decomposition \cite{Kata2} can push $T_\mathrm{B}$ to values considerably larger than those predicted for single 
particle coherent rotation (up to a factor 5). 
Thus we estimate $N_\mathrm{B}$ in the range of 250 magnetic
ions. Therefore, an observation of a hysteresis requires the existence of  regions where 250 CoV
ions interacting at 3$^\mathrm{rd}$ NN exist at concentrations larger than $\sim6\%$. Indeed this is a rather modest requirement.
\begin{figure}[ht]
\epsfxsize=9.0cm
\centerline{\epsffile{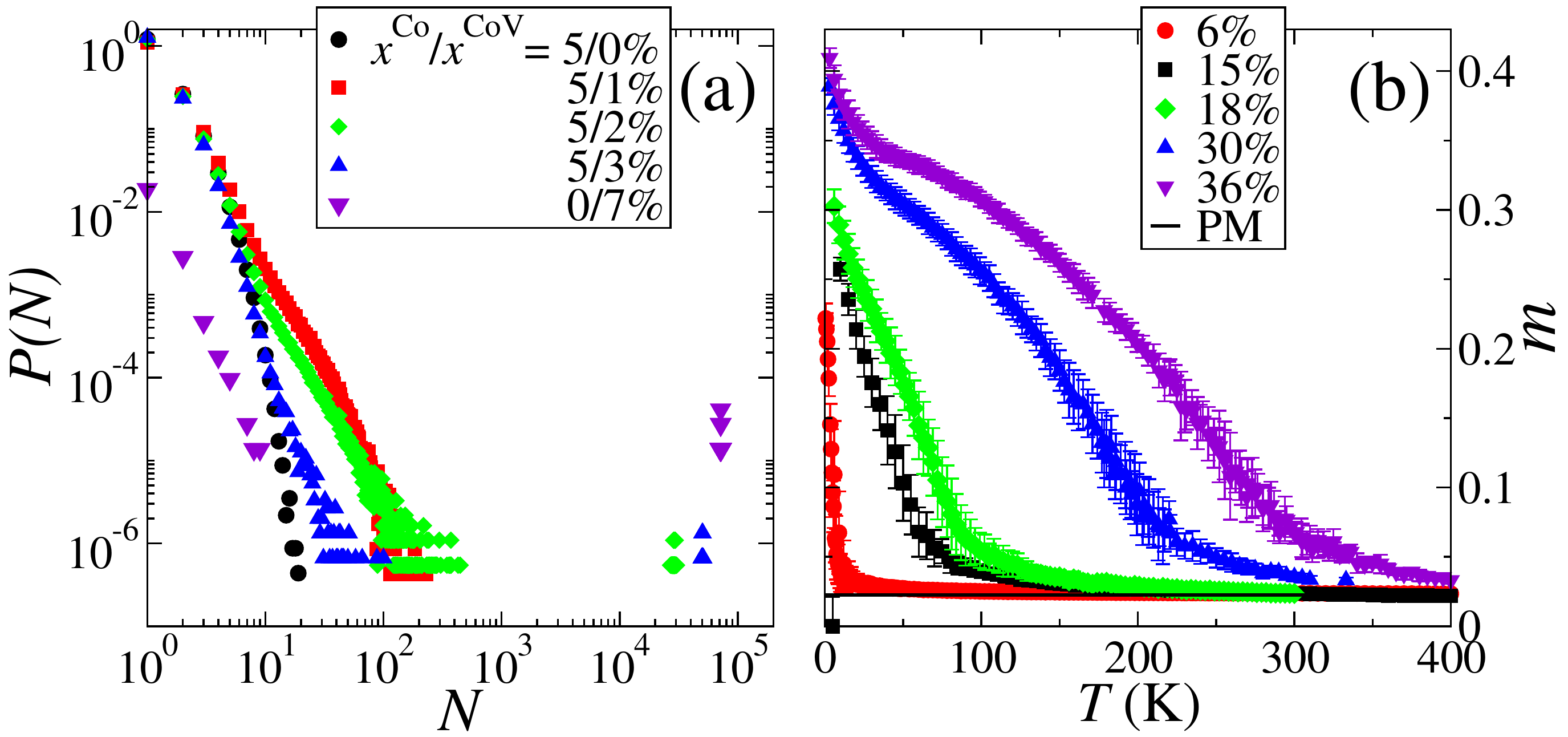}}
\centerline{\epsffile{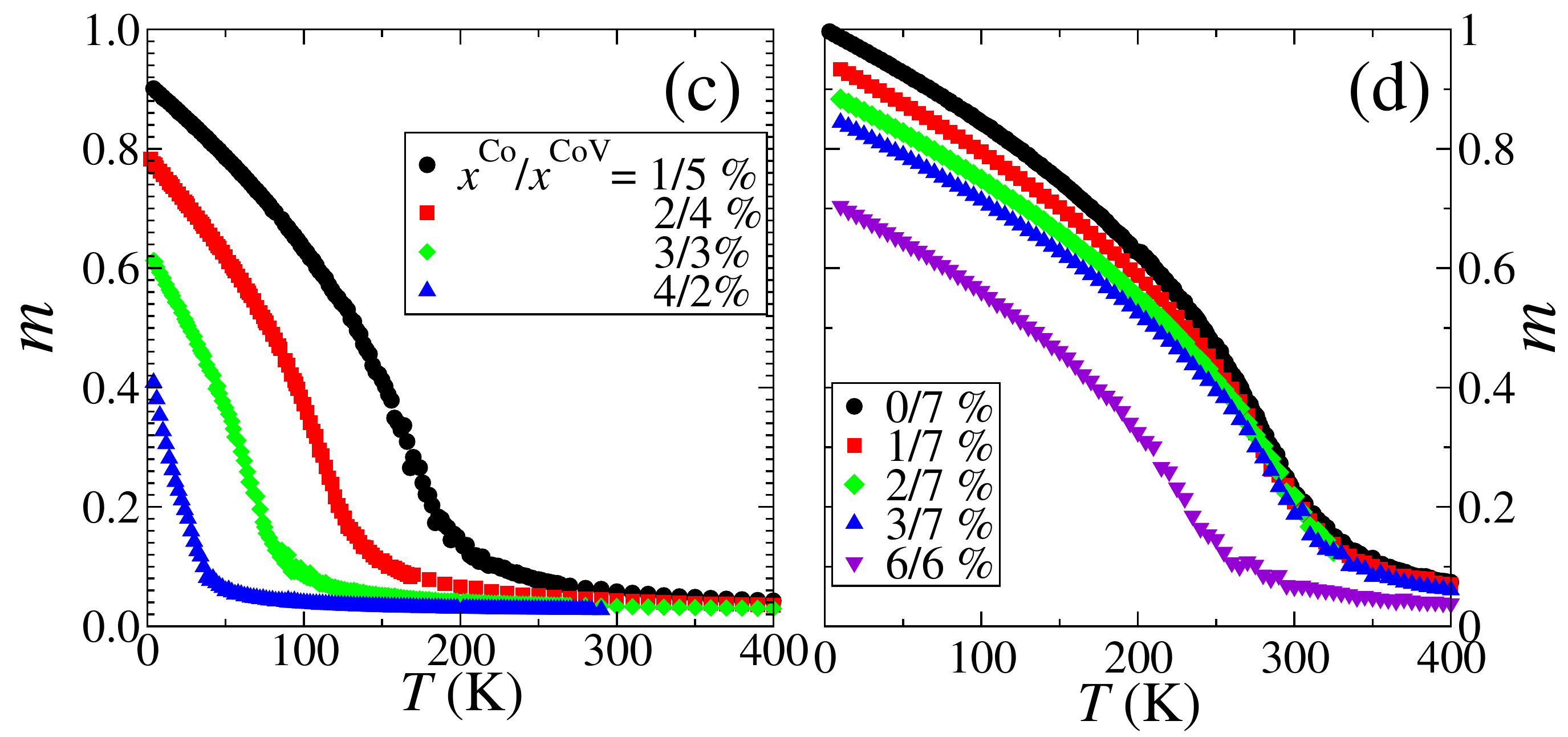}}
\caption{Monte Carlo simulations for the two center model: (a) cluster distribution $P(N)$ as a function of the number $N$ of magnetic ions 
in the cluster for different $x^\mathrm{Co}$ and $x^\mathrm{CoV}$ ($x^\mathrm{Co}$/$x^\mathrm{CoV}$); (b) magnetization curves 
$m(T)$ at different Co concentration for $x^\mathrm{Co}/x^\mathrm{CoV}=5$; (c) $m(T)$ for a total Co concentration of 6\% and
different $x^\mathrm{Co}/x^\mathrm{CoV}$; (d) $m(T)$ for $x^\mathrm{CoV}$ above percolation (7\%) and various $x^\mathrm{Co}$
($x^\mathrm{Co}$/$x^\mathrm{CoV}$).}
\label{Fig4}
\end{figure}

In Fig.~\ref{Fig4}a we present typical cluster distributions $P(N)$ as a function of the cluster size $N$ for various
concentrations of Co$^{2+}$ and CoV. These have been obtained by filling randomly a wurtzite lattice comprising
$10^6$ sites. As expected $P(N)$ moves from small to large clusters as $x^\mathrm{CoV}$ is increased with respect to 
$x^\mathrm{Co}$. In particular one notes that already for $x^\mathrm{CoV}=2\%$ large clusters
appear in the distribution, which becomes bi-modal at $x^\mathrm{CoV}>6\%$, i.e. above $x_\mathrm{c}$. 
We emphasize that these $P(N)$ have been obtained from a completely random distribution, i.e. neglecting
the tendency to clustering suggested by the pairing energy.

Finally we investigate the thermodynamical properties of our two center model by performing Monte Carlo simulations
for a Heisenberg energy ($|\mathbf{S}_{i}|=1$) 
\begin{equation}
{\cal{H}}_\mathrm{eff}=\sum_{\langle i,j\rangle}^\mathrm{Co,CoV}J_{ij}\mathbf{S}_i\cdot\mathbf{S}_{j}\:.
+\sum_{i}^\mathrm{Co,CoV}D(\mathbf{S}_i\cdot\hat{n})^2\;.
\end{equation}
The exchange parameters are chosen to mimic the short and long range exchange between Co$^{2+}$ and CoV respectively. 
At NN $J_{ij}$ is AFM for Co$^{2+}$ pairs (15~meV) and FM for CoV pairs (50~meV) and between Co$^{2+}$ and CoV
(50~meV). Moreover it extends to 2$^\mathrm{nd}$ (15~meV) and 3$^\mathrm{rd}$ NN (5~meV) for CoV pairs. The last term accounts 
for an hard-axis easy-plane anisotropy ($|\hat{n}|=1$) \cite{sati06}.
Note that we implicitly assume doping in the CoV impurity band and neglect the NN FM
component of the exchange between Co$^{2+}$ pairs. Given the uncertainty over the precise microscopic configuration and
the relative abundance of the various complexes (Table \ref{Tab1}) our numerical values are only representative
and certainly conservative. However, even with this choice $T_\mathrm{C}=250$~K above percolation 
($x^\mathrm{CoV}=7\%$) suggesting that RTF is indeed possible.

In Fig.~\ref{Fig4}b we present the reduced magnetization $m=M/M_\mathrm{s}$ as a function of $T$ for different Co concentrations, while
keeping $x^\mathrm{Co}/x^\mathrm{CoV}=5$. The magnetization curves show a transition from a concave upwards shape
at low concentrations to a convex one for high. As the concentration increases one
encounters the two percolation thresholds, respectively for CoV and for Co. This produces the change in shape, which however is
complete only above 20\%, i.e. when the Co$^{2+}$'s start to percolate. Interestingly $m$ never reaches 1 because of the strong 
NN AFM interaction among Co. 

We also investigate the interplay between Co$^{2+}$ and CoV. In figure Fig.~\ref{Fig4}c $m(T)$ is calculated by keeping the total Co 
concentration to 6\% and by varing the ratio between $x^\mathrm{Co}$ and $x^\mathrm{CoV}$, while in Fig.~\ref{Fig4}d we keep 
$x^\mathrm{CoV}=7$\% (above percolation) and change $x^\mathrm{Co}$. In both pictures $x^\mathrm{Co}$ is well below
the percolation threshold for NN and the high-temperature region of $m(T)$ is almost entirely dominated by $x^\mathrm{CoV}$.
For instance one may note that for $x^\mathrm{CoV}=5$\% $m(T)$ approaches zero roughly at the same $T$ regardless
of whether $x^\mathrm{Co}$ is 1\% (Fig.~\ref{Fig4}c) or 25\% (Fig.~\ref{Fig4}b for a total concentration of 30\%). In contrast the 
amount of Co$^{2+}$ affects the low temperature region, where the strong AFM interaction can drastically alter
the total magnetic moment. For instance $m(0)$ drops from 1 to about 0.8 if $x^\mathrm{Co}$ is increased from 1 to 3\%, by 
keeping $x^\mathrm{CoV}=7$\%. 

\begin{figure}[ht]
\epsfxsize=8.5cm
\centerline{\epsffile{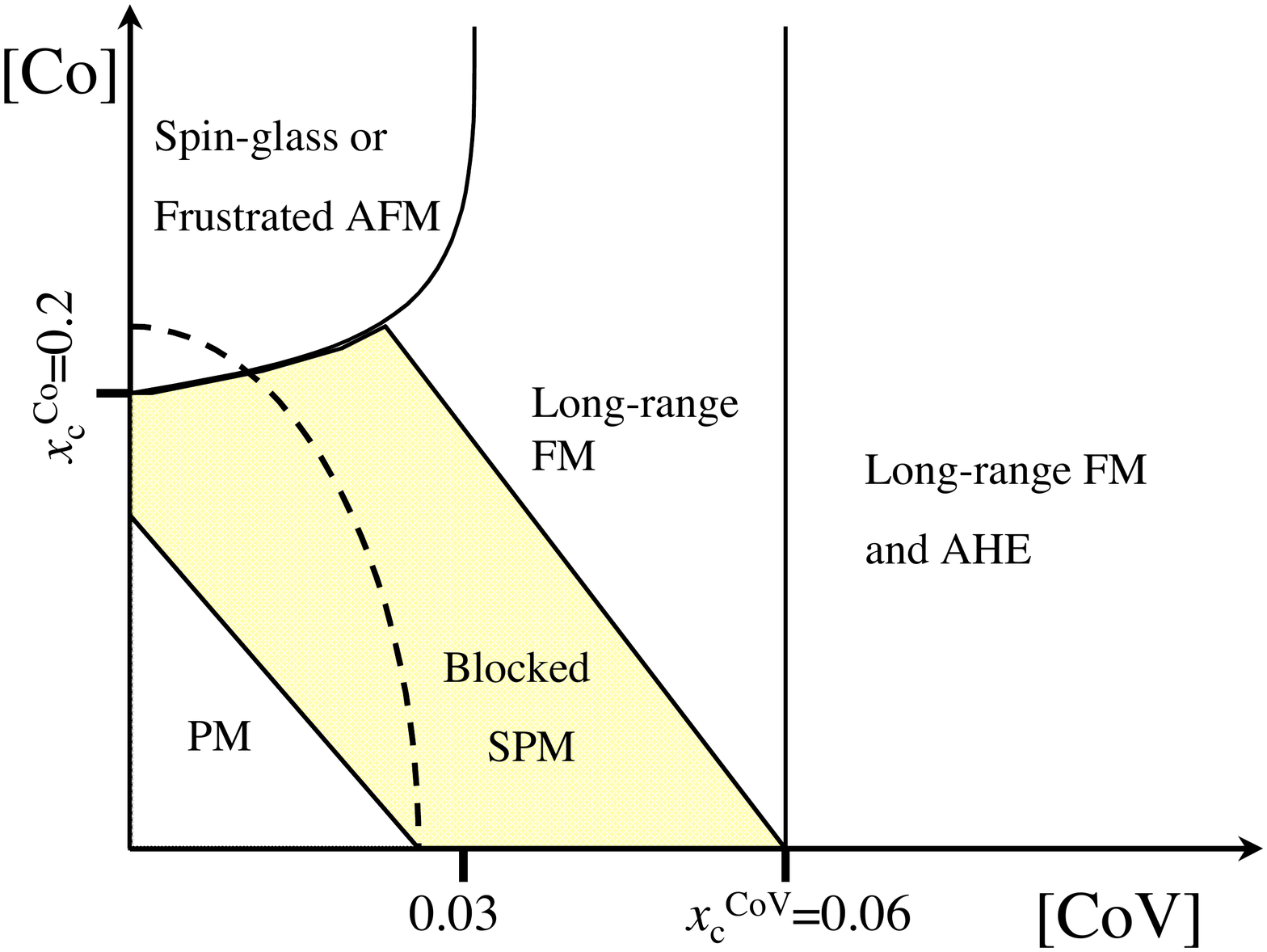}}
\caption{Proposed phase diagram for (Zn,Co)O as a function of the relative concentrations of Co and CoV. The yellow area is
the blocked superparamagnetic (SPM) region, where both magnetic moment and hysteresis can be detected. The dashed line
at small $x$ delimits the region where most of the experiments are conducted. Finally the FM region is partition into two regions 
depending on whether or not anomalous Hall effect can be detected.}
\label{Fig5}
\end{figure}

We are now in a position to propose a phase diagram for (Zn,Co)O based on the relative concentration of Co$^{2+}$
and CoV (Fig.~\ref{Fig5}). The most important feature is the presence of what we called a blocked superparamagnetic phase.
This is below $x_\mathrm{c}^\mathrm{CoV}$ and $x_\mathrm{c}^\mathrm{Co}$, but nevertheless allows one the detection 
of both a net magnetic moment and hysteresis at room-temperature. For larger $x$ a long-range FM ground state 
emerges, which however is limited by the short range AFM interaction of the Co ions. Therefore we predict either a frustrated antiferromagnet or a spin-glass for $x^\mathrm{Co}>x_\mathrm{c}^\mathrm{Co}$ and $x^\mathrm{CoV}\ll x^\mathrm{Co}$. 
We emphasize that bulk measurements (hysteresis or magnetization) can hardly distinguish between the FM and the 
blocked superparamagnetic phase, and more local probes are needed. In particular further insights would be provided by 
a thorough analysis of small angle neutron scattering data similar to the case of disordered ferromagnets \cite{Beni}.
Alternatives may be muon rotation, high resolution EPR and energy-dispersive X-ray spectroscopy.

Finally we partition the long-range FM region into two regions separated by the CoV percolation threshold 
$x_\mathrm{c}^\mathrm{CoV}$. 
For $x^\mathrm{CoV}>x_\mathrm{c}^\mathrm{CoV}$ percolation among CoV is achieved and one expects measurable conductivity from the 
impurity band. Since the exchange is strong an anomalous Hall effect (AHE) should be detected. This is not expected for 
ferromagnetism below $x_\mathrm{c}^\mathrm{CoV}$ since the conductivity is then dominated by band conductivity
which is weakly affected by Co$^{2+}$, given the small exchange. Note that  this phase diagram says little about the overall conductivity 
of (Zn,Co)O, which in turn can be determined by electrons in the conduction band. 
Moreover, the precise location of the phase boundaries depends on
details such as the concentration of electron donors. In the
extreme case of fully compensated samples the blocked superparamagnetic phase may
even disappear entirely.

In conclusion, by using a combination of DFT and Monte Carlo simulations, we have 
demonstrated that the observed RTF in (Zn,Co)O can be attributed to blocked superparamagnetism. This
develops at concentrations below those required by long-range ferromagnetism. However, even this
model requires a second magnetic dopant in addition to Co$^{2+}$ substituting Zn, capable of mediating magnetic interaction
beyond nearest neighbours. We have identified the Co-V$_\mathrm{O}$ pair as the most likely candidate and demonstrated that
such center can indeed promote long range coupling, if additional $n$ doping is present. These findings draw a new roadmap
for designing diluted magnetic oxides, where the engineering of intrinsic defects play the leading role. For instance 
paramagnetic samples can be turned ferromagnetic by prolonged exposure to Ti vapours, which produce high
concentrations of V$_\mathrm{O}$ \cite{ZnODefect}.

\section*{Methods}

Density functional theory calculations have been performed by using our newly developed approximate version of the self-interaction
correction scheme (ASIC) \cite{DasSIC}. This is implemented numerically in the localized basis set code {\it Siesta} \cite{Siesta}.
Here the ASIC scaling parameter $\alpha$ was set to 1/2, which is the value appropriate for describing the
electronic structure of mid- to wide-gap semiconductors \cite{DasSIC}.
In all the simulations we have considered unit cells ranging from 128 to 256 atoms. The basis set used was as follows:
Zn: DZ-$s$, DZ-$p$, SZ-$d$, O:  DZ-$s$, DZ-$p$, SZ-$d$, Co: DZ-$s$, DZP-$p$, DZ-$d$ (SZ=single zeta, DZ=double zeta,
DZP=double zeta polarized \cite{Siesta}). The grid cutoff (equivalent to plane wave-cutoff) was 
650~Ry and we have considered 18 $k$-points in the full Brillouin zone for the 128 atom cell and appropriate scaling for other cells. 
Standard conjugate gradient geometrical relaxation was performed until the forces were smaller than 0.04~eV/\AA.

Monte Carlo simulations were performed with the Metropolis algorithm as implemented in a home-made package. 
Typical simulation cells for the two center Heisenberg model contain between 1,600 to 5,000 magnetic ions and we always use
periodic boundary conditions. The systems are equilibrated until an initially AFM and FM replica have 
converged to the same value (typically 10,000 steps) and then the Monte Carlo measurements are
taken by sampling 60,000 new steps. Disorder averages are taken over 32 different samples at each concentration. $T_\mathrm{C}$
for the case of $x^\mathrm{CoV}=7\%$ is evaluated from the specific heat (see supplementary materials).


\small{

}

\section*{Acknowledgment}

This work is sponsored by Science Foundation of Ireland under the
grants SFI02/IN1/I175 and 05/IN1/I853. We thank TCHPC and ICHEC for
providing computational support. 

\end{document}